\documentclass[british]{article}
\usepackage[T1]{fontenc}
\usepackage[latin9]{inputenc}
\usepackage{geometry}
\usepackage{url}
\usepackage{amstext}
\usepackage{graphicx}

\makeatletter
\newcommand{\lyxaddress}[1]{
\par {\raggedright #1
\vspace{1.4em}
\noindent\par}
}

\date{}
\makeatother

\usepackage{babel}
\begin{document}

\title{Wisdom of the institutional crowd}

\author{Kevin Primicerio $^{1,}$\thanks{\protect\url{kevin.primicerio@centralesupelec.fr}}
, Damien Challet $^{1,}$\thanks{\protect\url{damien.challet@centralesupelec.fr}}
, Stanislao Gualdi $^{1,2,}$\thanks{\protect\url{stanislao.gualdi@gmail.com}}}
\maketitle

\lyxaddress{$^{1}$ Laboratory of Mathematics in Interaction with Computer Science,
CentraleSup\'elec, Grande Voie des Vignes, 92290 Ch\^atenay-Malabry,
France\\
$^{2}$ Capital Fund Management, 23 rue de l\textquoteright Universit\'e,
75007, Paris, France}
\begin{abstract}
The average portfolio structure of institutional investors is shown
to have properties which account for transaction costs in an optimal
way. This implies that financial institutions unknowingly display
collective rationality, or Wisdom of the Crowd. Individual deviations
from the rational benchmark are ample, which illustrates that system-wide
rationality does not need nearly rational individuals. Finally we
discuss the importance of accounting for constraints when assessing
the presence of Wisdom of the Crowd.
\end{abstract}

\section{Introduction}

The collective ability of a crowd to accurately estimate an unknown
quantity is known as the ``Wisdom of the Crowd'' \cite{surowiecki2005wisdom}
(WoC thereafter). In many situations, the median estimate of a group
of unrelated individuals is surprisingly close to the true value,
sometimes significantly better than those of experts \cite{galton1907vox,hill2011expert,landemore2012many,nofer2014crowds}.
WoC may only hold under some conditions \cite{surowiecki2005wisdom,davis2014crowd}:
for example social imitation is detrimental as herding may significantly
bias the collective estimate \cite{lorenz2011social,muchnik2013social}.
WoC is a reminiscent of collective rationality without explicit individual
rationality: when it applies, it is a consistent aggregation of possibly
inconsistent individual estimates \cite{hogarth1978note}. This is
to be contrasted with the mainstream economic paradigm which takes
a short-cut by assuming that collective rationality reflects individual
rationality, where only a ``typical'' decision maker \textendash{}
the representative agent \textendash{} is considered \cite{hartley2002representative}
or team reasoning where the individual agents explicitly optimize
the collective welfare \cite{colman2008collective}. Aggregation of
quite diverse individual actions, especially in a dynamic context
where expectations are continuously revised, is still an open problem
\cite{kirman1992whom}.

Although almost all known examples of WoC are about a single number
or coordinate, there is no reason why WoC could not be found for whole
functional relationships between several quantities. For example,
Haerdle and Kirman analyse the prices and volume of many transactions
in Marseille fish market: while the relationship between these two
quantities is rather noisy, the market self-organises so that when
more fish are sold, prices are lower, as revealed by a local average
\cite{Haerdle1995}. More generically, many simple relationships found
in Economics textbooks may only hold on average, but not for each
agent or each transaction. 

Asset price efficiency is an obvious instance of WoC in Finance: it
states that current prices, determined by the actions of many traders,
are the best possible estimates and fully reflect all available information
\cite{malkiel1970efficient,malkiel2003efficient,fama1998market}.
Another WoC candidate is portfolios. While many market participants,
especially investment funds, strive to build optimal portfolios, each
following its own criteria and constraints (performance objective,
risk, tracking error, etc.), the question here is whether their collective
behaviour may be related to a rational benchmark. Fortunately, this
implies that we do not need to understand the minute details of all
the portfolios and can focus on average quantities instead.

\section{Wisdom of crowd}

Let us define some necessary quantities to be more precise. At time
$t,$ fund $i$ has capital $W_{i}(t)$ which is invested into $n_{i}(t)$
securities among $M(t)$ existing ones. As a result, each security
$\alpha$, whose capitalization is denoted by $C_{\alpha}(t)$, is
found in $m_{\alpha}(t)$ portfolios. The explicit time dependence
is dropped hereafter. 

The only quantity defined above which depends on asset allocation
strategies of fund $i$ is $n_{i}$, the number of securities it chooses
to invest in. Our main hypothesis is thus that WoC is found in the
average relationship between $n_{i}$ and $W_{i}$. A simple rational
benchmark is proposed by \cite{de2010turnover} : when a fund with
capital $W_{i}$ is able to invest the same amount in each of the
$n_{i}$ chosen securities and if the transaction cost does not depend
on the security, then the optimal $n_{i}$ is such that
\begin{equation}
W_{i}\propto n_{i}^{\mu}.\label{eq:Wi_vs_ni}
\end{equation}

where the exponent $\mu$ is determined by the transaction costs fee
structure; for example, proportional transaction costs lead to $\mu=1$,
while a fixed cost per transaction corresponds to $\mu=2$ (see \cite{de2010turnover}
for more details). Allowing for individual fluctuations, Eq.~(\ref{eq:Wi_vs_ni})
becomes $\log W_{i}=\mu\log n_{i}+\epsilon_{i}$, where $\epsilon_{i}$
has zero average. Denoting local average of $x_{i}$ by $x$, the
local average of Eq.~(\ref{eq:Wi_vs_ni}) yields
\begin{equation}
W\propto n^{\mu}.\label{eq:W_vs_n}
\end{equation}

Flat fee per transaction ($\mu=2$) is a popular request of large
clients of broker. \cite{de2010turnover} find indeed that for wealthy
individual investors and asset managers, exponent $\mu=2$ within
statistical uncertainty. We will thus test the occurrence of WoC from
the value of exponent $\mu$. More precisely, our hypothesis is that
if (i) the effective transaction cost per transaction is the same
for all assets and (ii) funds are able to build equally weighted portfolios,
then Eq.~(\ref{eq:W_vs_n}) holds and that $\mu=2$, which is a sign
of WoC.

Both conditions must cease to hold for larger investment funds. Indeed,
condition (i) cannot be true for them since large trades (even when
split into meta-orders) have a price impact which grows with their
size and depends on volatility and average turnover \cite{bouchaud2010priceimpact}.
Condition (ii) ceases to hold for large funds which spread their investments
on many securities: because the capitalization of assets and their
average daily turnover are very heterogeneous, large funds cannot
invest enough money in assets with a small capitalization so as to
build an equally-weighted portfolio. As a result, on average, the
local average $W$ is expected to increase more slowly as a function
of $n$ in the large $n$ region; equivalently, the exponent $\mu$
is expected to be smaller than $2$. In summary, two different regimes
should emerge: one with $\mu_{<}=2$ for small enough $n$ and $\mu_{>}<2$
for larger $n$. 

\begin{figure}
\begin{centering}
\includegraphics[scale=0.12]{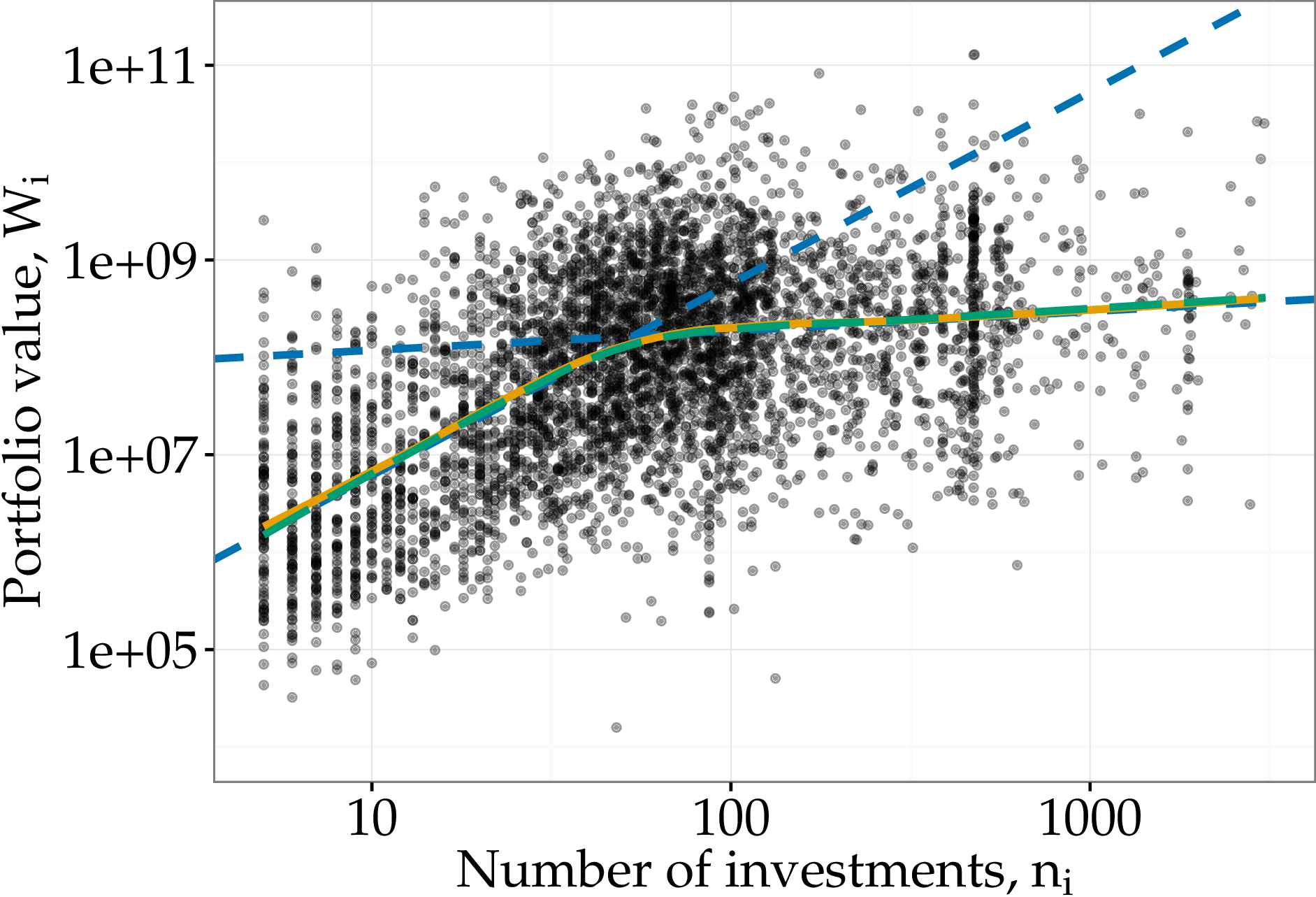}
\par\end{centering}
\caption{Total mark-to-market value $W_{i}$ as a function of the number of
investments $n_{i}$, with a robust locally weighted regression fit
(yellow line) and two linear fits (blue dashed lines) for two different
ranges of n. Robust locally weighted regression fit for the simulated
data (in green). \label{fig:wi_vs_ni_emp}}
\end{figure}

Figure~\ref{fig:wi_vs_ni_emp} plots $W_{i}$ versus $n_{i}$ in
logarithmic scale: a cloud of point emerges, with a roughly increasing
trend. The large amount of noise confirms the great diversity of fund
allocation strategies. WoC may only appear in some average behaviour.
This is why we computed a locally weighted polynomial regression \cite{cleveland1992local}.
As expected, two distinct regions appear. In each of them, the local
regression follows a roughly linear behaviour. 

\begin{figure}
\begin{centering}
\includegraphics[scale=0.12]{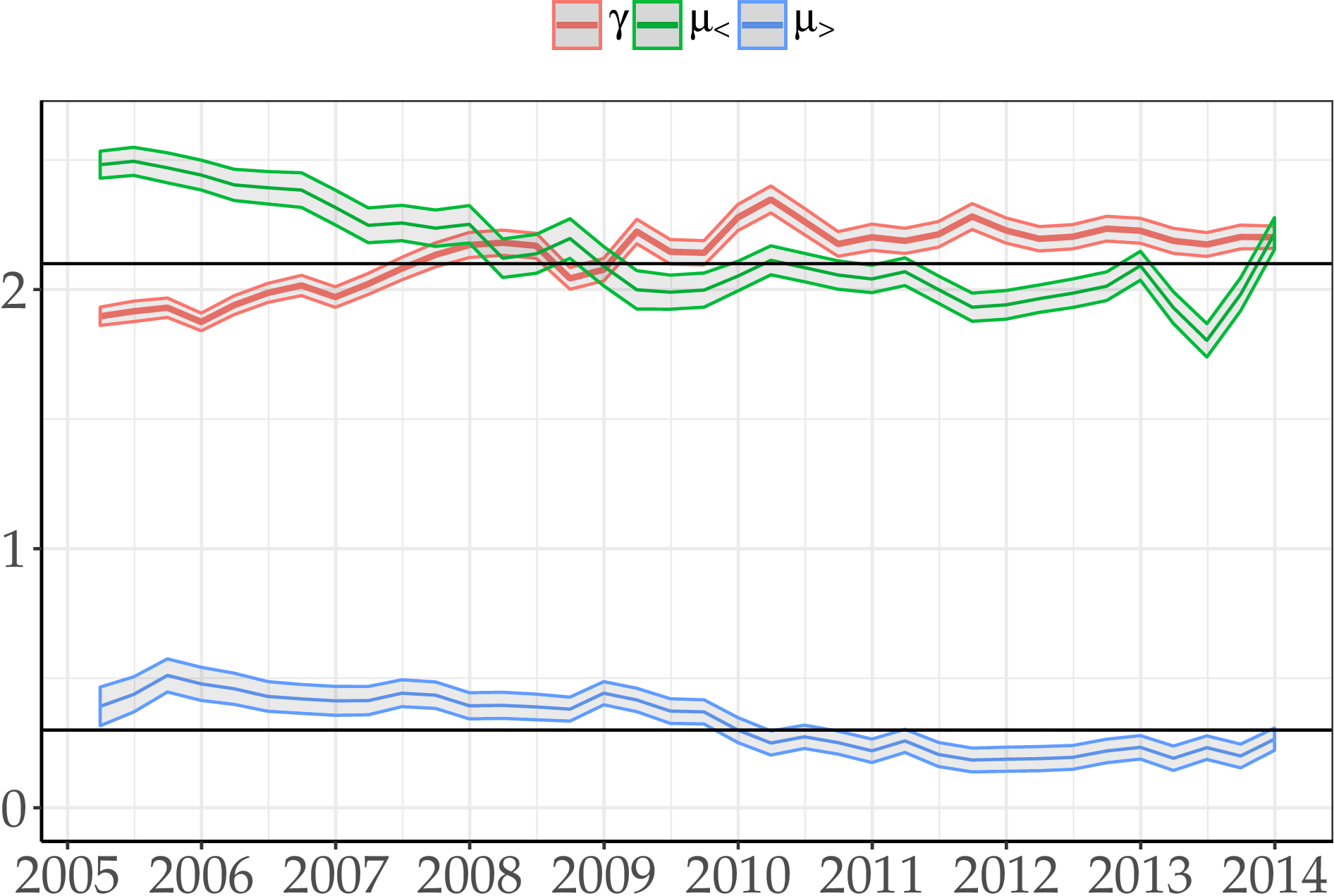}
\par\end{centering}
\begin{centering}
\includegraphics[scale=0.12]{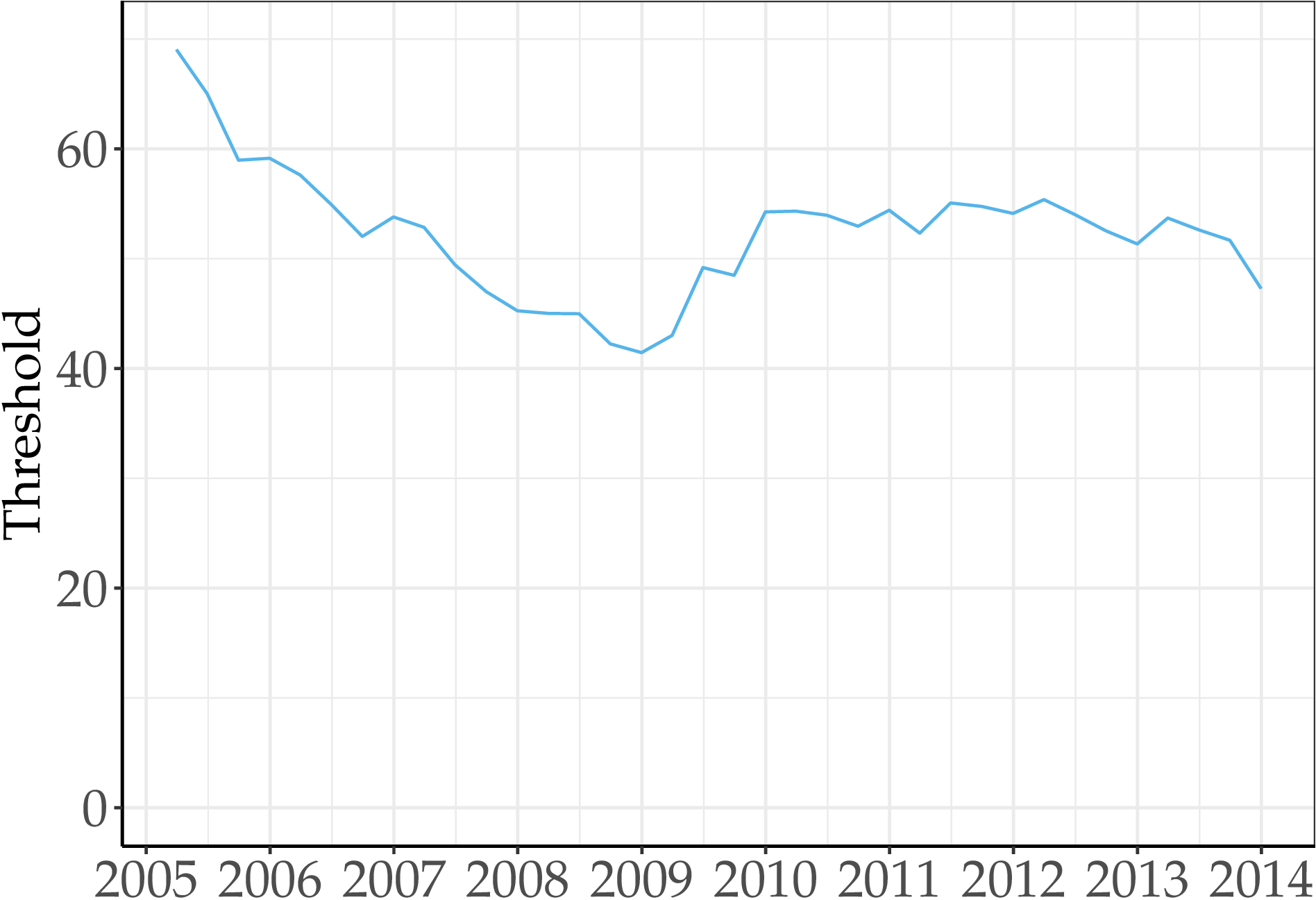}
\par\end{centering}
\caption{Top: Temporal evolution of the coefficients $\mu_{<}$, $\mu_{>}$
and $\gamma$. Bottom: Temporal evolution of the value of the cross-over
point $n^{*}$ between the two regions as a function of time. It reaches
a global minimum after the 2008 crisis. \label{fig:time_evolution}}
\end{figure}

The cross-over point $n^{*}$ between the two regions is algorithmically
determined for each quarterly snapshot (see S.I.); it is stable as
time goes on (see Fig.~\ref{fig:time_evolution} in S.I.). The two
exponents $\mu_{<}$ and $\mu_{>}$ are quite stable as a function
of time as well (see Fig. \ref{fig:time_evolution} in S.I.); their
time-averages $\overline{\mu_{<}}\simeq2.1\pm0.2$ and $\overline{\mu_{>}}\simeq0.3\pm0.1$
are markedly different, which points to distinct collective ways of
building portfolios in these two regions. 

\begin{figure}
\begin{centering}
\includegraphics[scale=0.33]{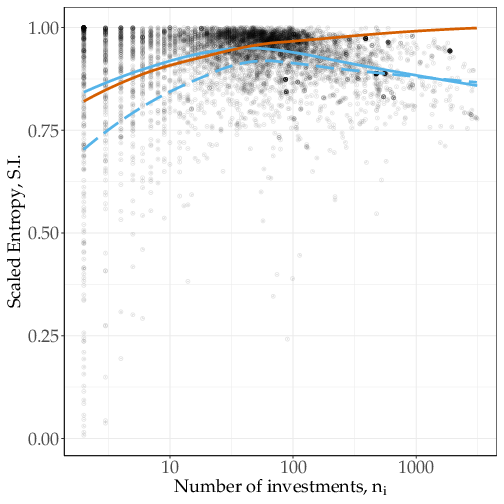}
\par\end{centering}
\centering{}\caption{Scaled Shannon entropy $S_{i}$ as a function of the number of investments
$n_{i}$ for all the funds on 2013-03-31 (circles) and robust local
weighted regression fit, for all positions (blue line), numerical
simulation of the effect of price fluctuations on the entropy on initially
equally weighted portfolios (red line) , where a volatility similar
to the observed volatility in the real data, is applied; robust local
weighted regression fit restricted to the unchanged portfolio positions
from the previous time step, multiplied by the ratio between the simulated
entropy for the full portfolio and the restricted portfolio (dashed
blue line); \label{fig:shannon-entropy} }
\end{figure}

So far, $\overline{\mu_{<}}$ is compatible with the WoC hypothesis.
Let us check the validity of conditions (i) and (ii) above. When condition
(ii) is not satisfied, then condition (i) also must cease to hold,
thus we can focus on the former. Condition (ii) says that the diversity
of investment fractions $p_{i\alpha}=W_{i\alpha}/W_{i}$ for $W_{i\alpha}>0$
must be very small among $\alpha$. This may be summarized in a single
number by the scaled Shannon Entropy $S_{i}=-\frac{1}{\text{log}_{2}n_{i}}\sum_{\alpha}p_{i\alpha}\log_{2}p_{i\alpha}$,
which equals 1 and is maximal when all the non-null $p_{i\alpha}$
are equal. Figure.~\ref{fig:shannon-entropy} reports the scaled
entropy $S_{i}$ of all the funds for a given time snapshot, together
with the local average $S$. The latter increases up to about $n\simeq n^{*}$
and then decreases. The fact that $S<1$ is due in part to price fluctuations:
even if fund $i$ builds an equally weighted portfolio at time $t$
(thus $S_{i,t}=1$), $S_{i,t+1}<1$ at a later date. The importance
of this mechanism is confirmed by Monte-Carlo simulations: the red
line of Fig.~\ref{fig:shannon-entropy} shows the effect of natural
asset price evolution on perfectly equally weighted portfolios after
three months, using asset price volatility measured in our dataset
between the time of the snapshot and the three previous months: the
resulting scaled entropy $S_{MC}$ increases as a function of $n$,
mirroring the local average of $S_{i}$ in the same figure for $n<n^{*}$.
Thus, the decrease for $n>n^{*}$ is due to impossibility for larger
funds to build equally-weighted portfolios. A further argument supporting
our claim that investment funds strive to build equally weighted portfolios
(on average) is provided by the entropy measured on the set of common
positions between two consecutive snapshots multiplied by $S_{MC}(n_{i})/S_{MC}(n_{i,restricted})$
in order to account for the dependence of $S$ on $n$; the local
average of the resulting entropy corresponds to the dashed blue line:
it is clearly smaller than the entropy of the new portfolio, hence
new positions purposefully bring $S_{i}$ closer to equally weighted
portfolios. Therefore, condition (ii) is valid when $\mu=2$; conversely,
$\mu\ne2$ when condition (ii) ceases to hold.

Quite tellingly, the same exponent was found for large private investors
and asset managers (with much smaller amounts of money under management).
Thus the collective behaviour of large investment funds is essentially
the same one. Since one finds the same exponent $\mu$ over many decades
of portfolio values for a wide spectrum of market participants, and
since $\mu=2$ corresponds to a realistic transaction cost per transaction,
we argue that WoC is a plausible explanation of the average portfolio
structure. Note that $\mu=2$ does not imply that funds really face
constant transaction cost per transaction, only that their population
acts as if it does. Finally, we stress that WoC holds for a whole
functional relationship over many decades of $n$ and $W$, not only
for a single number, which considerably extends its reach.

\section{Asset selection model}

So far, bringing to light WoC in the $\mu=2$ region only required
to focus on the number of securities in a portfolio, not on how funds
select securities. This implicitly assumed that funds could invest
in all securities they wished, which is clearly not the case in the
large diversification region: the fact that the exponent $\mu$ is
much smaller in this region implies that funds need on average to
split their investments into many more securities. This is most likely
due to liquidity constraints: large funds cannot invest as much as
they wish in some assets because there are simply not enough shares
to build a position larger than a certain size without impacting too
much their prices. Each fund has its own way to determine the maximal
amount to invest in a given security $\alpha$; a common criterion
is to limit the fraction $W_{i\alpha}/C_{\alpha}$. Fig.~\ref{fig:fmax}
in S.I. strongly suggests that each fund fixes its upper bound 
\begin{equation}
f_{i}^{(\text{max})}\ge\max_{\alpha}f_{i\alpha}\,\,\,\,\text{where }f_{i\alpha}=\frac{W_{i\alpha}}{C_{\alpha}}.\label{eq:fi}
\end{equation}
 It turns out that $f_{i}^{\text{max}}$ is highly heterogeneous among
funds $\log{}_{10}\left(f_{i}^{\text{max}}\right)\simeq-3.0\pm1.0$
(see Fig.~\ref{fig:fimax_density}), which reflects both the heterogeneous
ways of portfolio construction and also the confidence of a fund in
its abilities to execute large trades without too much price impact.
The existence of such limits implies that portfolios are less likely
to be equally weighted in the large diversification region, as seen
indeed in the decrease of the average portfolio weights scaled entropy
for $n\geq70$ (blue line in Fig.~\ref{fig:shannon-entropy}). 

\begin{figure}
\begin{centering}
\includegraphics[scale=0.12]{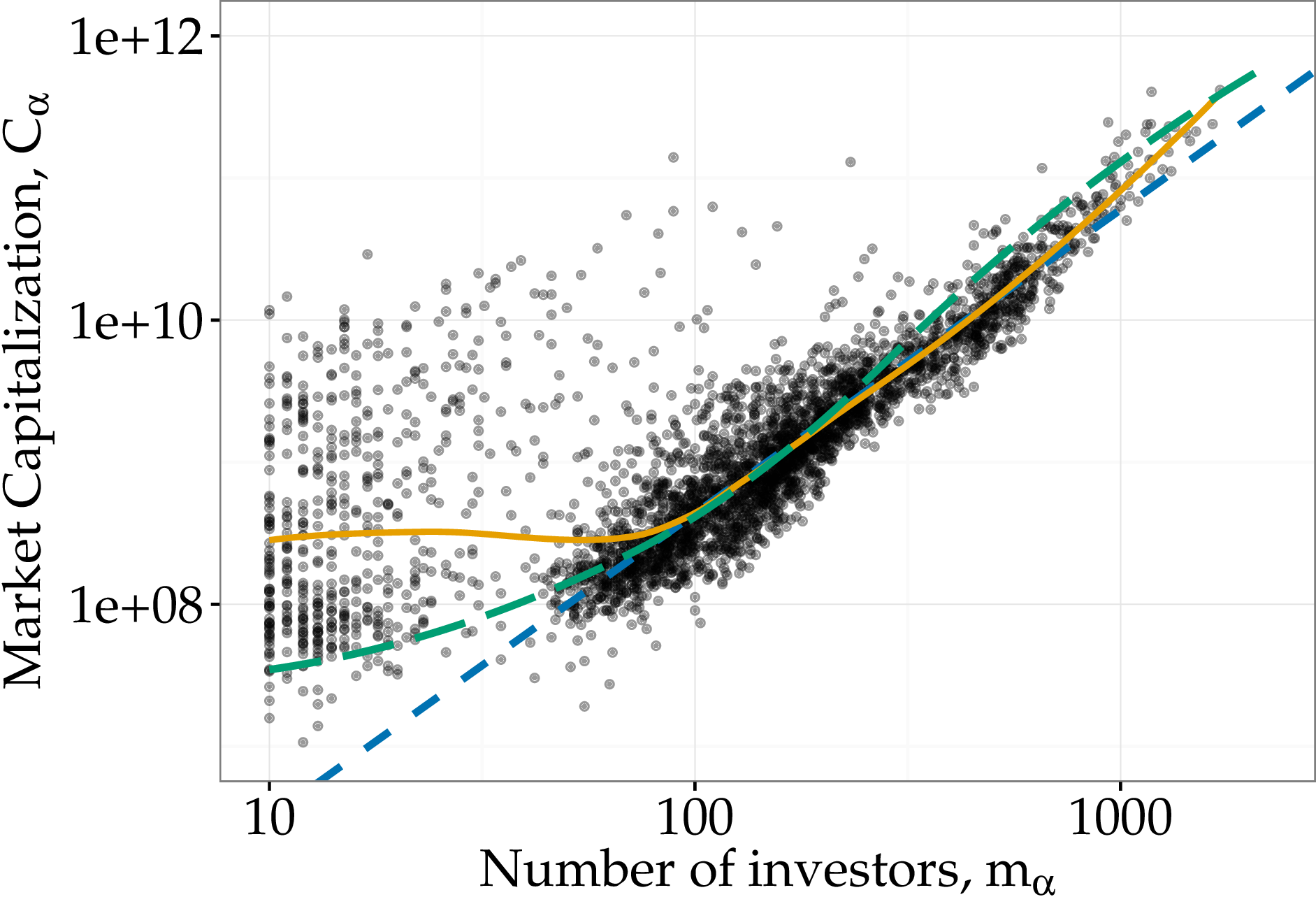}
\par\end{centering}
\caption{Market capitalization of securities as a function of the number of
investors in logarithmic scale. From the local non-linear robust fit
(yellow line) we observe a linear relationship for assets with more
than about 100 investors. The blue dashed line corresponds to a linear
fit on that group of asset. Hence $W_{\alpha}\propto m_{\alpha}^{\text{\ensuremath{\gamma}}}$,
with $\gamma\simeq2.1$. Robust locally weighted regression fit for
the simulated data (in green). \label{fig:ca_vs_na}}
\end{figure}

Funds, however, do not invest in a randomly chosen security, even
in the low diversification region. Figure \ref{fig:ca_vs_na} displays
a scatter plot of the capitalization $C_{\alpha}$ of each security
$\alpha$ versus $m_{\alpha}$, the number of funds which have invested
in this security, together with a local non-linear fit. Similarly
to $W$ vs $n$, one finds a power-law relationship 
\begin{equation}
\log C_{\alpha}=\gamma\log m_{\alpha}+\epsilon_{\alpha}\label{eq:Ca_vs_na}
\end{equation}
 for large enough $m$ (see S.I.). Hence in local average notations,
$C\propto m^{\gamma}$. Exponent $\gamma$ is stable during the period
2007-2014 (see Fig. \ref{fig:time_evolution} in S.I.) and its average
$\bar{\gamma}\simeq2.2\pm0.1$.

\begin{figure}
\begin{centering}
\includegraphics[scale=0.12]{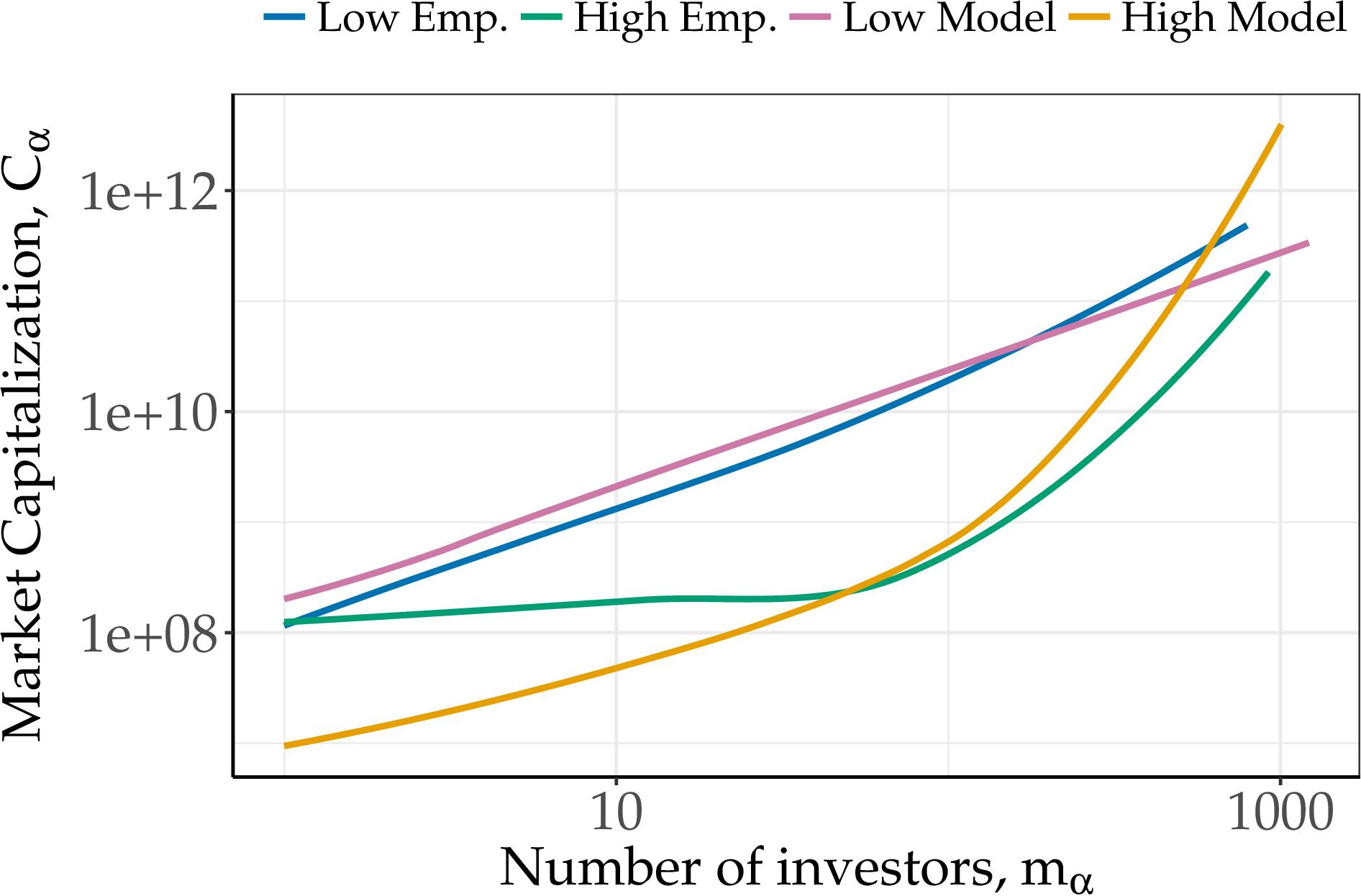}
\par\end{centering}
\caption{We separate the contribution from the low and highly diversified region.
The origin of the discrepancy observed in Fig.~\ref{fig:ca_vs_na}
appears to be mainly due to the highly diversified region (Green dots
for the empirical data, and orange dots for the model). \label{fig:ca_vs_na_separation}}
\end{figure}

In short, one needs to introduce a model of how funds choose to invest
in securities to reproduce the average behaviour of both Eqs (\ref{eq:Ca_vs_na})
and (\ref{eq:Wi_vs_ni}). Since one sees a cross-over between two
types of behaviour rather than an abrupt change, we create logarithmic
bins of the axis $n_{i}$ and denote the bin number of fund $i$ by
$[n_{i}]$. Two mechanisms must be specified: how a fund selects security
$\alpha$ and how much it invests in it. The latter point is dictated
by Fig.~\ref{fig:fmax} in the large $n_{i}$ region where fund $i$
invests $W_{i\alpha}=f_{i}^{(\text{max})}C_{\alpha}$; for the sake
of simplicity, we approximate $f_{i}^{(\text{max})}$ by the median
value of $f_{i}^{(\text{max})}$ in the bin $[n_{i}]$, denoted by
$f_{[n_{i}]}^{(\text{max})}$. In the small diversification region,
we assume that $n_{i}=n_{i}^{\text{opt}}$, thus $W_{i\alpha}=W_{i}/n_{i}^{\text{opt}}$
to be consistent with our previous results. We choose a security selection
mechanism that rests on the market capitalization $C_{\alpha}$ of
a security $\alpha$ (see S.I.) which is a good proxy of the liquidity
(Fig.~\ref{fig:ca_vs_wia}). We perform Monte-Carlo simulations from
the empirical selection probabilities and $f_{[n_{i}]}^{(\text{max})}$
and display the resulting $W$ vs $n$ and $C$ vs $m$ in Figs \ref{fig:wi_vs_ni_emp}
and \ref{fig:ca_vs_na} (continuous green lines), in good agreement
with the local averages (continuous orange lines). One notices a discrepancy
in the relationship $C$ vs $m$ for large $n$, which mainly comes
from funds in the large diversification region. (See Fig~\ref{fig:ca_vs_na_separation}
S.I).

The large diversification region illustrates how constraints may considerably
modify the rational benchmark. While the above mechanism of security
selection is able to reproduce adequately the behaviour of well diversified
funds, we could not find a rational benchmark for the dependence of
$f^{\text{max}}$ and $n_{i}$. Thus, the case for WoC in the large
diversification region is not entirely closed.

\section*{Data}

Our dataset consists of an aggregation of the following publicly available
reports (in order of reliability): the SEC Form 13F, the SEC's EDGAR
system forms N-Q and N-CSR and (occasionally) the form 485BPOS. Our
work focuses on the period starting from the first quarter of 2005
to the last quarter of 2013.

These forms are filled manually and are thus error prone. We partially
solve this issue by cross-checking different sources (which often
contains overlapping information) and by filtering data before processing
(see details in S.I.).

The main limitation of this dataset is that it provides accurate figures
for long positions only. The other positions (short, bonds, ...) are
most of the time only partially known. The frequency of the dataset
is also inhomogeneous: data for most of the funds are quarterly updated
(depending on regulations), hence we decided to restrict ourselves
to 4 points in a year only. Such frequency is probably too low for
investigating the dynamics of individual behaviour but is not a problem
for we focus on an aggregate and static representation of the investment
structure.

\section*{Discussion and conclusion}

While WoC is commonly applied to a population collectively guessing
a single number, we investigate here a fundamentally different situation
and provide evidence for a collective functional optimization of the
asset ownership structure. What the reference function should be is
dictated by optimality arguments. In the case of financial markets,
the rational benchmark was not related to the efficient market hypothesis,
but to the way a large population of professional fund managers build
their portfolios. Whereas each fund has its own benchmark with respect
to which the fund performance may be assessed, this, fortunately,
has no discernible influence on the average structure of their portfolio.
In addition, WoC is often meant as a collective guessing of non-experts;
one thus may conclude that the population investigated here has decidedly
more expertise than the subjects of other WoC studies. What kind of
expertise the typical fund manager has is not obvious, at least when
one looks at their pure performance (see e.g. \cite{barras2010false}).
In addition, the optimal relationship between the number of assets
in a portfolio and the value of the latter is clearly not broadly
known in these circles, as shown by the very large deviations from
the ideal case in Fig.~\ref{fig:wi_vs_ni_emp}, and the collective
expertise only appears when their decisions are suitably averaged.
The presence of WoC when the subjects face strong constraints, as
those of highly diversified funds, is more conjectural, and more work
will be needed in that respect.

At a higher level, our results suggest that, while individuals may
deviate much from the rational expectation theory, standard economic
theory may hold at a collective level, without need for micro-founded
individual decisions: the average decision may in some cases be approximated
by a rational, representative agent. Our results however only hold
on a snapshot of the system, for which individual fluctuations may
be averaged out. In a dynamic setting, the very large deviations from
the rational benchmark may not be neglected in the presence of feedback
loops \cite{gualdi2015tipping}. In other words, the dynamics of these
fluctuations are worth investigating in their own right.

\section*{Acknowledgements}

S. Gualdi acknowledges support of Labex Louis Bachelier (project number
ANR 11-LABX-0019)

\bibliographystyle{vancouver}
\bibliography{references}

\clearpage{}

\section*{Supporting Information (SI)}

\begin{figure}
\begin{centering}
\includegraphics[scale=0.12]{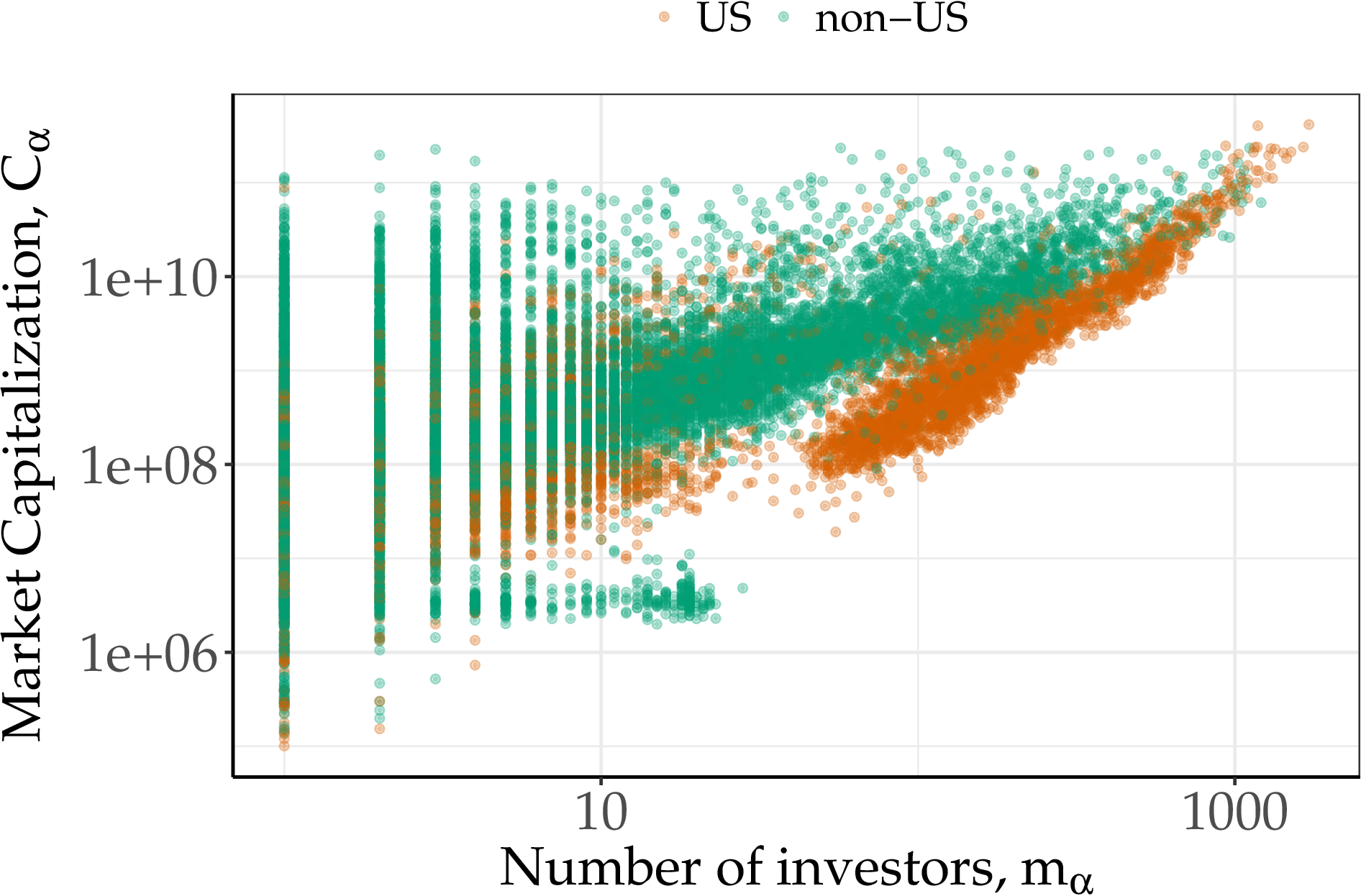}
\par\end{centering}
\begin{centering}
\includegraphics[scale=0.12]{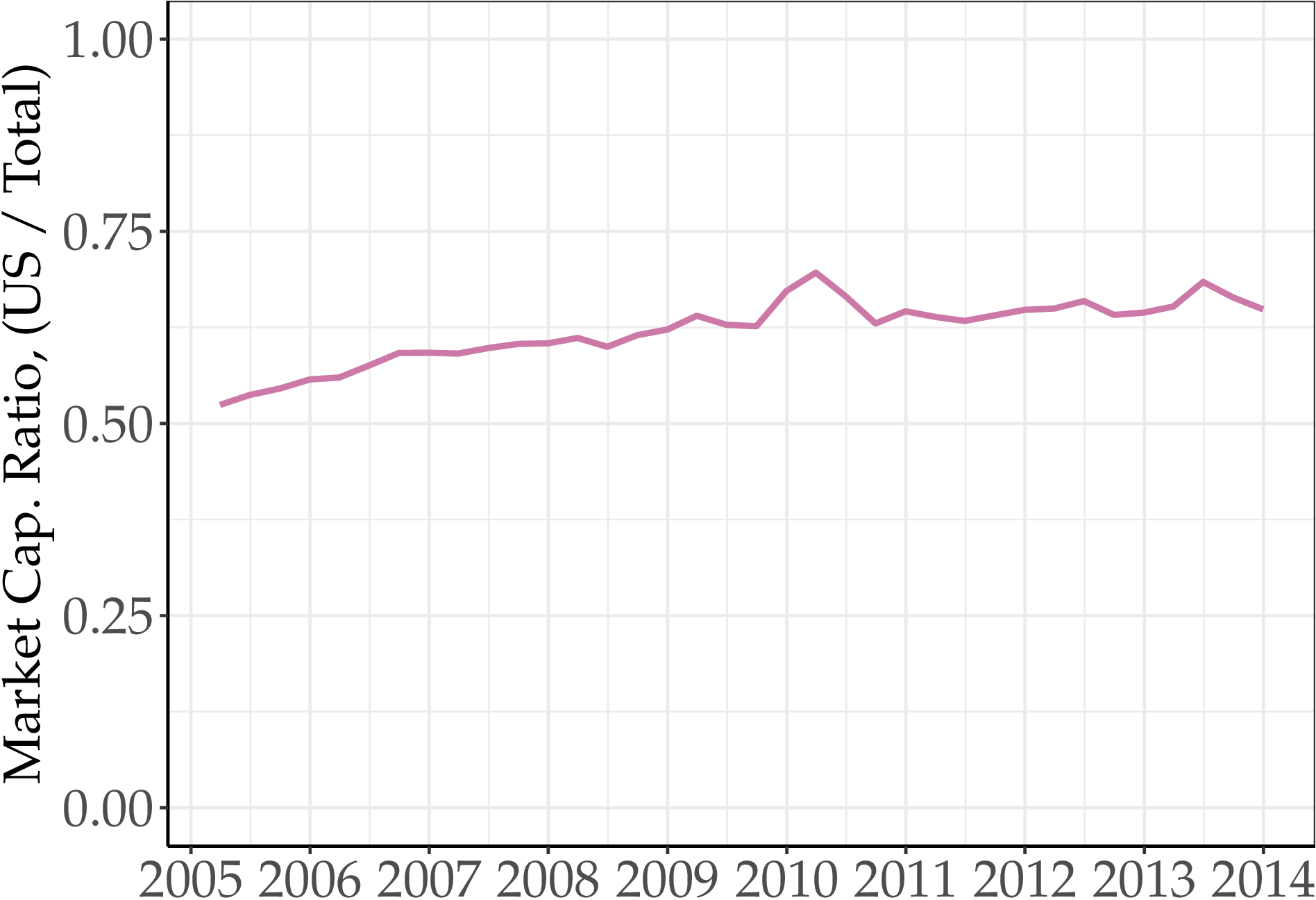}
\par\end{centering}
\caption{Top: Market capitalization as a function of the number of investors
for all securities. Bottom: Temporal evolution of the aggregated market
capitalization of US over the total market capitalization.\label{fig:ca_us_vs_other}}
\end{figure}

\section{Filtering}

In order to remove inconsistencies in the dataset, we applied the
following filters 

\subsection{Country of origin}

Our dataset is sparse and heterogeneous. Indeed, the quality of the
sources of data is directly related to each country's disclosure regulations.
For these reasons we decided to keep only the entities which use an
US based mail address.

About 60\% of the total market capitalization of the dataset is concentrated
in US based securities. Figure~\ref{fig:ca_us_vs_other} shows two
large clouds of dots, each of them corresponds to a different region
of origin: green (resp. orange) cloud corresponds to non-US (resp.
US) based securities. The origin of this large difference between
these two regions are not clear: it could for example come from differences
in regulations in non-US countries. It turns out that the ratio of
the investment values in US and non-US assets varies little as a function
of time (see Fig.~\ref{fig:ca_us_vs_other}), which does not affect
the exponent $\mu$ in Eq.~\ref{eq:Wi_vs_ni}. As a consequence we
focused on US securities.

\subsection{Frequency}

Large funds are requested to report their positions at a frequency
which depends on the applicable regulation. As a result, reporting
frequency ranges from monthly to yearly, most funds filing quarterly
reports. We therefore focused of the latter.

\subsection{Penny Stocks}

The ``penny stocks'', i.e., usually securities which trade below
\$5 per share in the USA, are not listed on a national exchange. Since
they are considered highly speculative investments and are subject
to different regulations, we filtered them out.

\subsection{Size}

We also filtered out small founds and securities and applied the following
filters: f$W_{i}>10^{5}$ USD, $C_{\alpha}>10^{5}$ USD, $n_{i}\geq5$,
$m_{\alpha}\geq10$. 

\subsection{Output}

\begin{figure}
\begin{centering}
\includegraphics[scale=0.12]{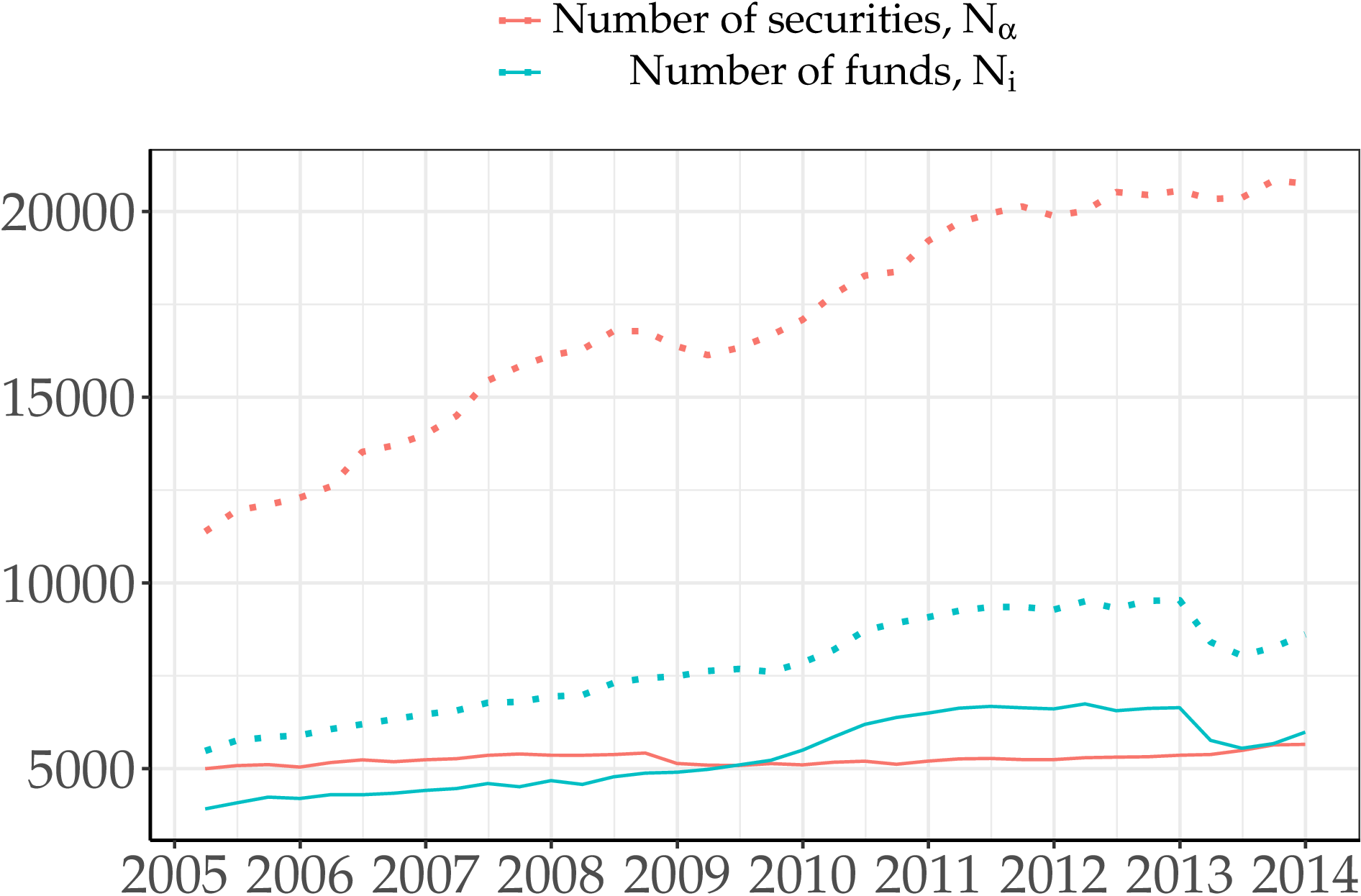}
\par\end{centering}
\caption{Temporal evolution of the number of funds $N_{i}$ and securities
$N_{\alpha}$ in the database. Unfiltered in dashed lines and US based
only in solid lines. \label{fig:evol_nb_entities}}
\end{figure}

We restricted our study to 36 quarterly snapshots starting from the
first quarter of 2005 and ending with the last quarter of 2013. Figure~\ref{fig:evol_nb_entities}
reports the evolution of the number of securities and funds in the
database before and after filtering.

\section{Asset selection modelling}

The framework we introduce in this paper follows a series of a few
elementary steps described below. The aim is for the model to be sensitive
to the different constraints which dominates the portfolio selection
of a fund.

\subsection{Finding $\mathbf{n^{*}}$}

For  date $t,$ we define the cross-over point $n^{*}$ between the
two regions which appear in the local polynomial regression. We determine
this point value with a likelihood maximization of the model 

\begin{equation}
W=\mu_{<}n+(\mu_{>}-\mu_{<})(n-n^{*})\theta(n-n^{*}),\label{eq:break-point}
\end{equation}

where $\theta(x)$ is the Heaviside function. We use a recursive method
to find parameters $\mu_{<}$, $\mu_{>}$ and $n^{*}$ \cite{muggeo2003estimating}.
Figure~\ref{fig:time_evolution} shows that $n^{*}$ is stable as
a function of time.

\subsection{Asset selection in the small diversification region $n_{i}<n^{*}$}

In this region, we consider the equally weighted portfolio hypothesis
to be true. Each position has a size $\frac{W_{i}}{n_{i}^{\textrm{opt }}}$,
where $n_{i}^{\text{opt}}$ is the optimal number of position computed
with eq~\ref{eq:Wi_vs_ni}. The funds select their asset randomly
with a probability proportional to $C_{\alpha}$. Also, in order to
construct an equally-weighted portfolio, a position is valid only
if it is of size $\frac{W_{i}}{n_{i}^{\textrm{opt}}}$. 

\subsection{Asset selection in the large diversification region $n_{i}\geq n^{*}$}

In this region, the liquidity constraints make it harder for funds
to keep an equally weighted portfolio and portfolio values are thus
spread on a larger number of assets. We propose here a stochastic
model of asset selection based on two main ingredients: first that
the selection probability of asset $\alpha$ by fund $i$ depends
on the diversification of a fund $n_{i}$ and on the scaled rank of
the capitalization of asset $\alpha$, and that the investment is
bounded by an hard constraint on the fraction of market capitalization
of asset $\alpha$.

We chose a security selection mechanism which rests on the scaled
rank of capitalization of security $\alpha$, defined as $\rho_{\alpha}=\frac{r_{\alpha}}{M}$
where $r_{\alpha}$ is the rank of capitalization $C_{\alpha}$ and
$M$ the number of securities at a given time. The selection probability
$P(W_{i\alpha}>0|\rho_{\alpha})$ is then obtained by parametric fit
to a beta distribution in each logarithmic bin. Note that we do not
use the same rank-based selection mechanism in the low-diversification
region because in this case it is harder to have a good fit with the
beta distribution. This is however only a minor point since the capitalization
is approximately power-law distributed and the two selection mechanisms
are basically equivalent (the rank is proportional to a power of the
capitalization) and indeed results are very similar in both cases.

\begin{figure}
\begin{centering}
\includegraphics[scale=0.12]{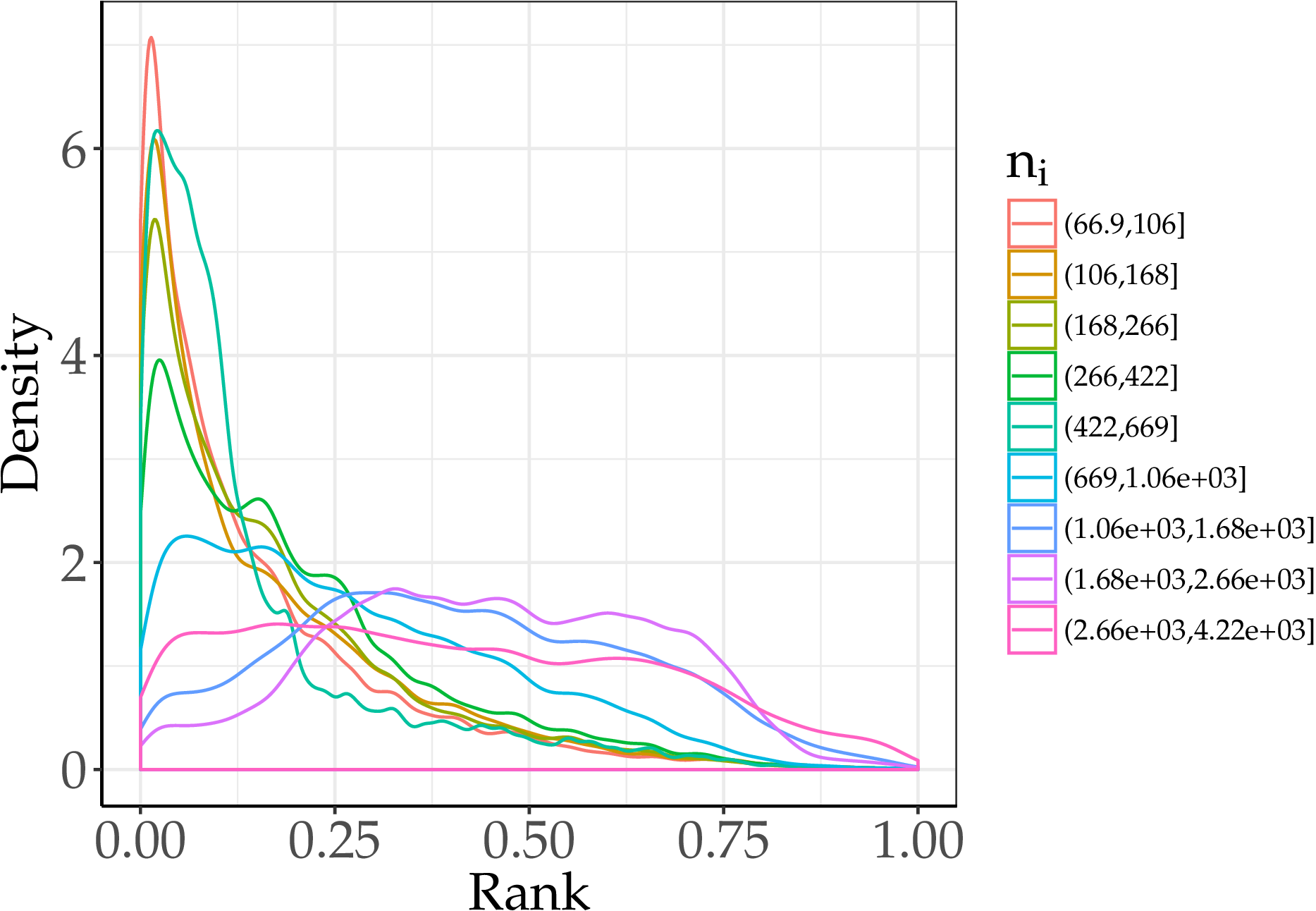}
\par\end{centering}
\begin{centering}
\includegraphics[scale=0.12]{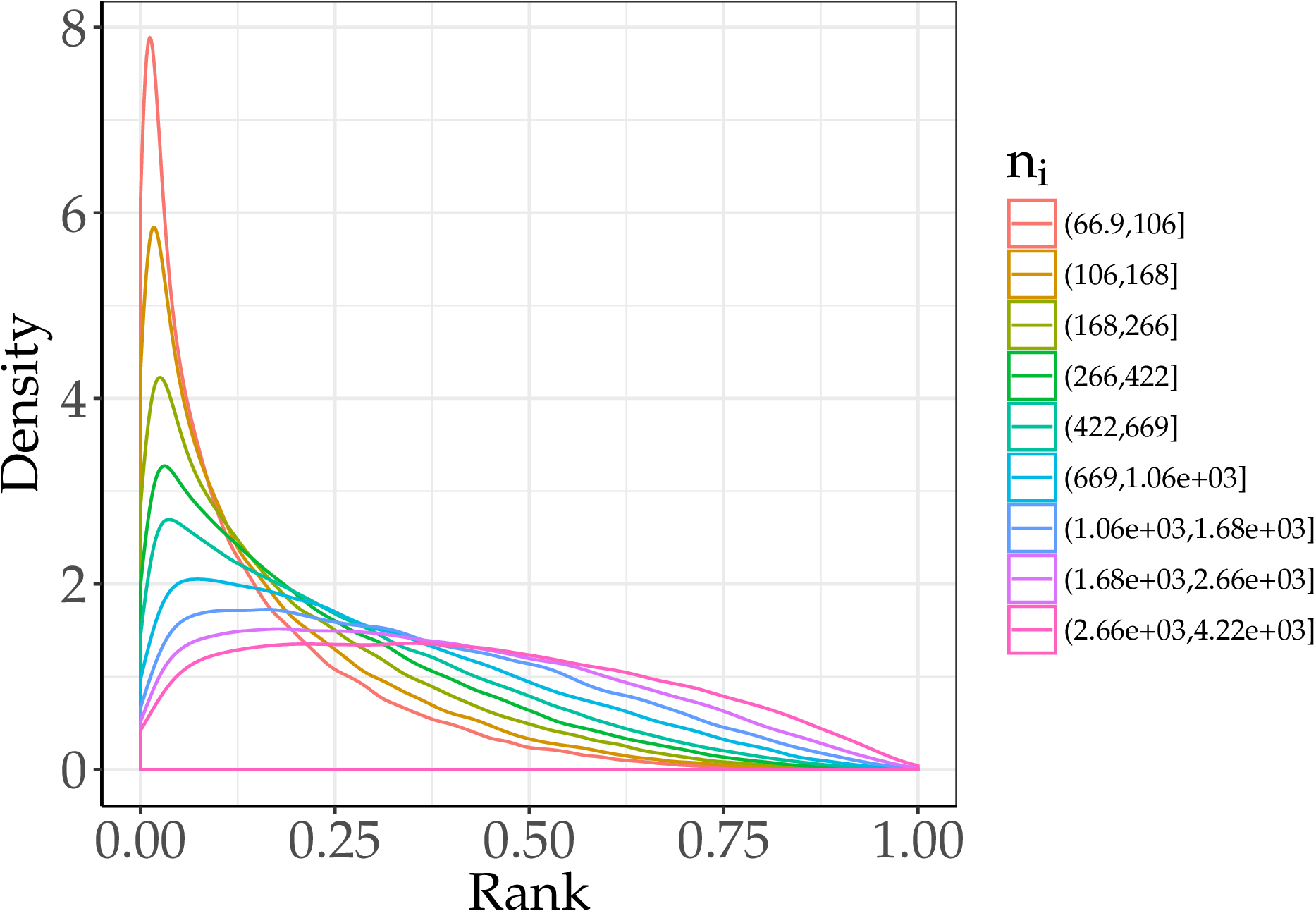}
\par\end{centering}
\caption{Top: Empirical probability density function of investing in a security
of scaled capitalization rank $\rho$ given the diversification $n_{i}$
of the fund. Bottom: Probability density function of investing in
a security of scaled capitalization rank $\rho$ given the diversification
$n_{i}$ of the fund, given by the model. \label{fig:asset_select}}
\end{figure}

Figure~\ref{fig:asset_select} shows that the distribution of the
ranks in which a fund is invested is sensitive to its diversification
$n_{i}$ for 2013-03-31. The Beta distribution, which is limited to
a $\left[0,1\right]$ interval, is flexible enough to describe the
asset selection mechanism of a fund.

\begin{equation}
f(x;a,b)=\frac{1}{B(a,b)}x^{a-1}(1-x)^{b-1},\label{eq:beta-dist}
\end{equation}

where $a$ and $b$ are the shape parameters of the distribution,
and $B$ is a normalization constant.

\begin{figure}
\begin{centering}
\includegraphics[scale=0.12]{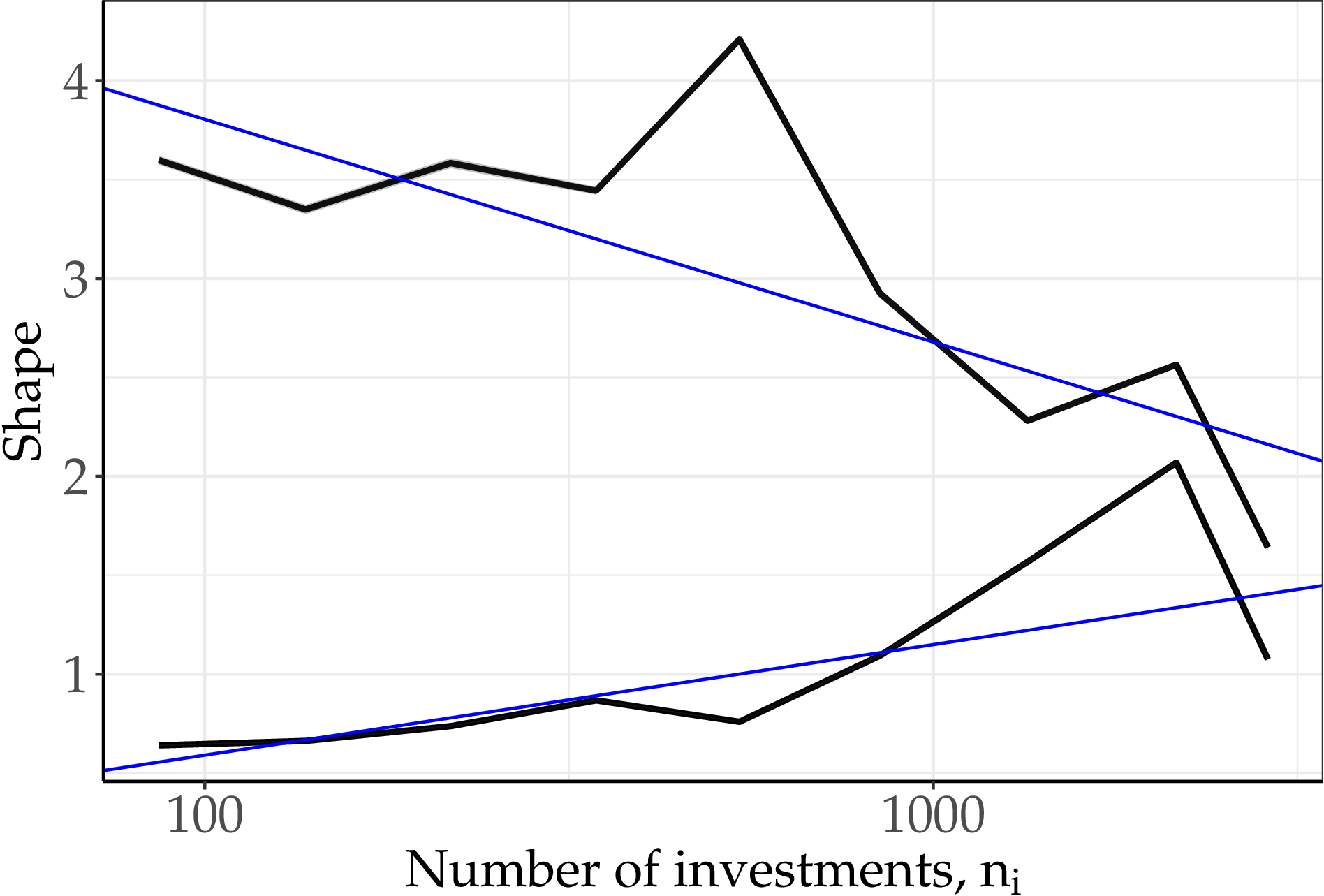}
\par\end{centering}
\caption{Coefficients $a$ and $b$ of the Beta Distribution \ref{eq:beta-dist}
as a function of $n_{i}$ \label{fig:beta_shape}. Linear fits are
for eye-guidance only.}
\end{figure}

\subsubsection{Maximum investment ratio}

\begin{figure}
\begin{centering}
\includegraphics[scale=0.12]{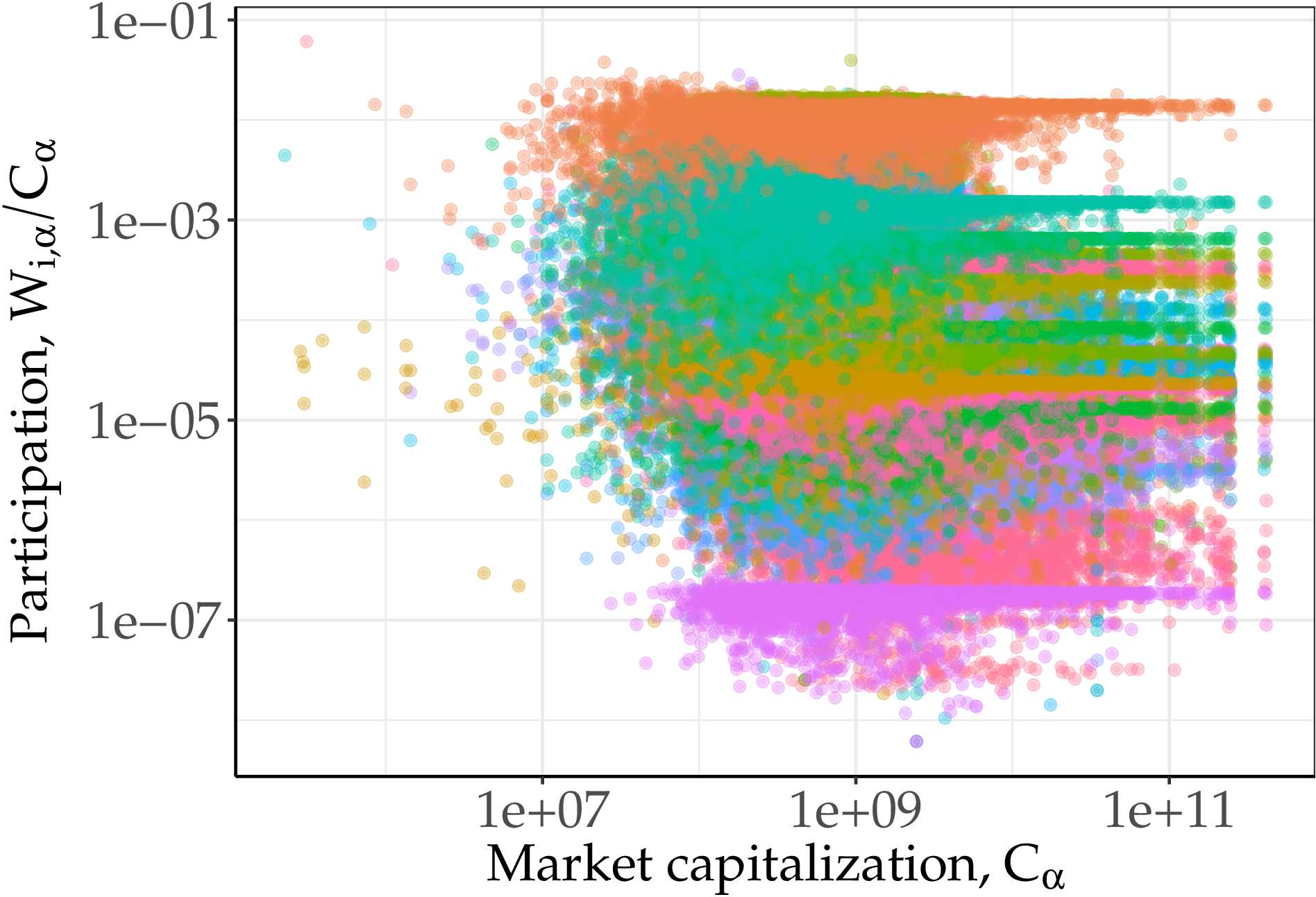}
\par\end{centering}
\begin{centering}
\includegraphics[scale=0.12]{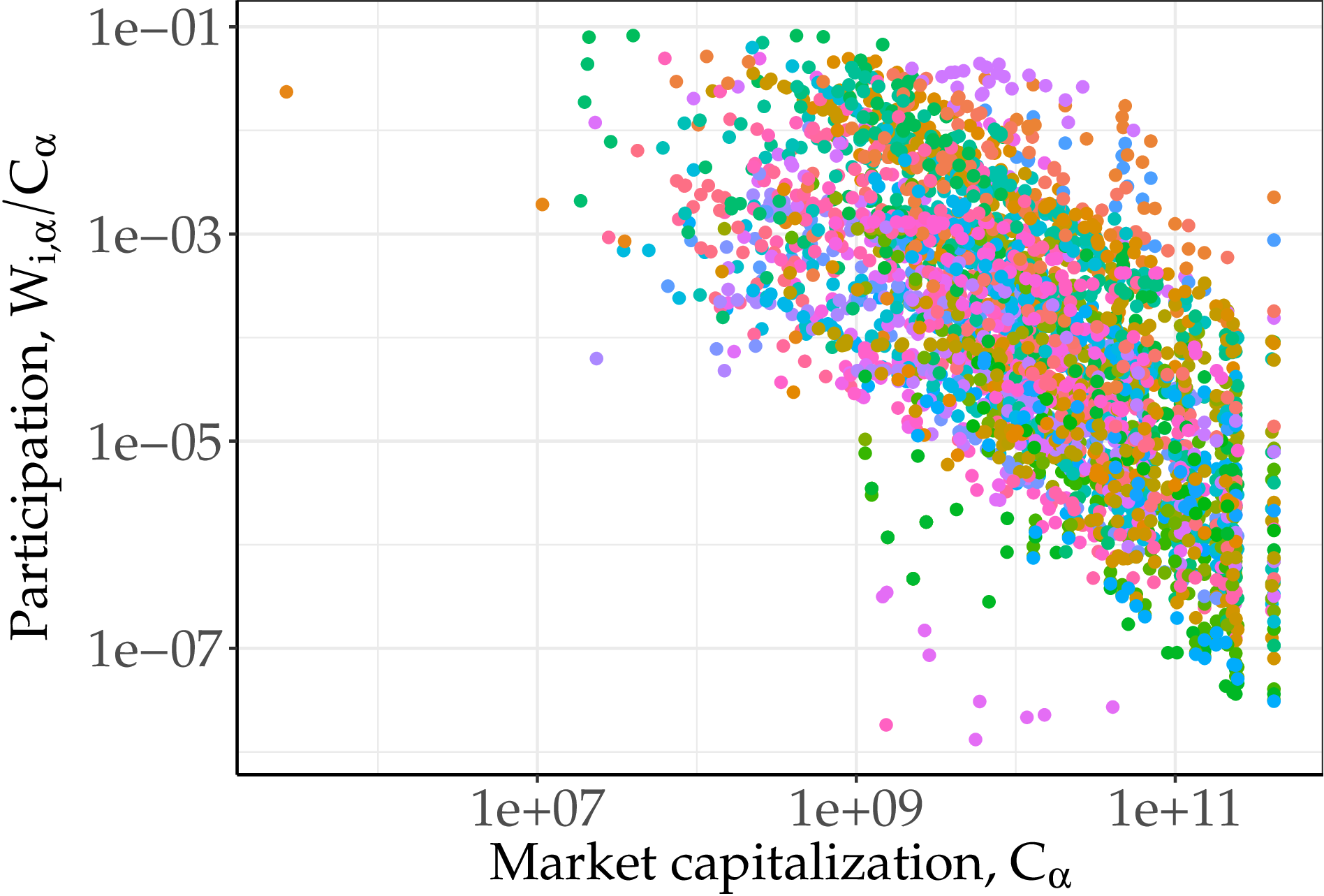}
\par\end{centering}
\caption{Fraction of the market capitalization of a security held by a fund.
Each color represent a different fund. Top: Funds with a large diversification
($n_{i}>800$). We can clearly see a delimitation for most of the
funds, which correspond to the maximum fraction $f_{i}^{\textrm{max}}.$
The value of $f_{i}^{\textrm{max}}$ widely differs from one fund
to another. Bottom: Funds with a low diversification ($n_{i}$<60),
$f_{i}^{\textrm{max}}$ doesn't appear. \label{fig:fmax}}
\end{figure}

The funds limit their investment in a given asset. They seem to follow
a simple rule: defining the investment ratio $f_{i,\alpha}=\frac{W_{i\alpha}}{C_{\alpha}}$,
one easily sees in Fig.~\ref{fig:fmax} that each fund has a maximum
investment ratio
\begin{equation}
f_{i}^{\textrm{max}}=\textrm{max}_{\alpha}\left(\frac{W_{i\alpha}}{C_{\alpha}}\right)\label{eq:fmax}
\end{equation}

\begin{figure}
\begin{centering}
\includegraphics[scale=0.12]{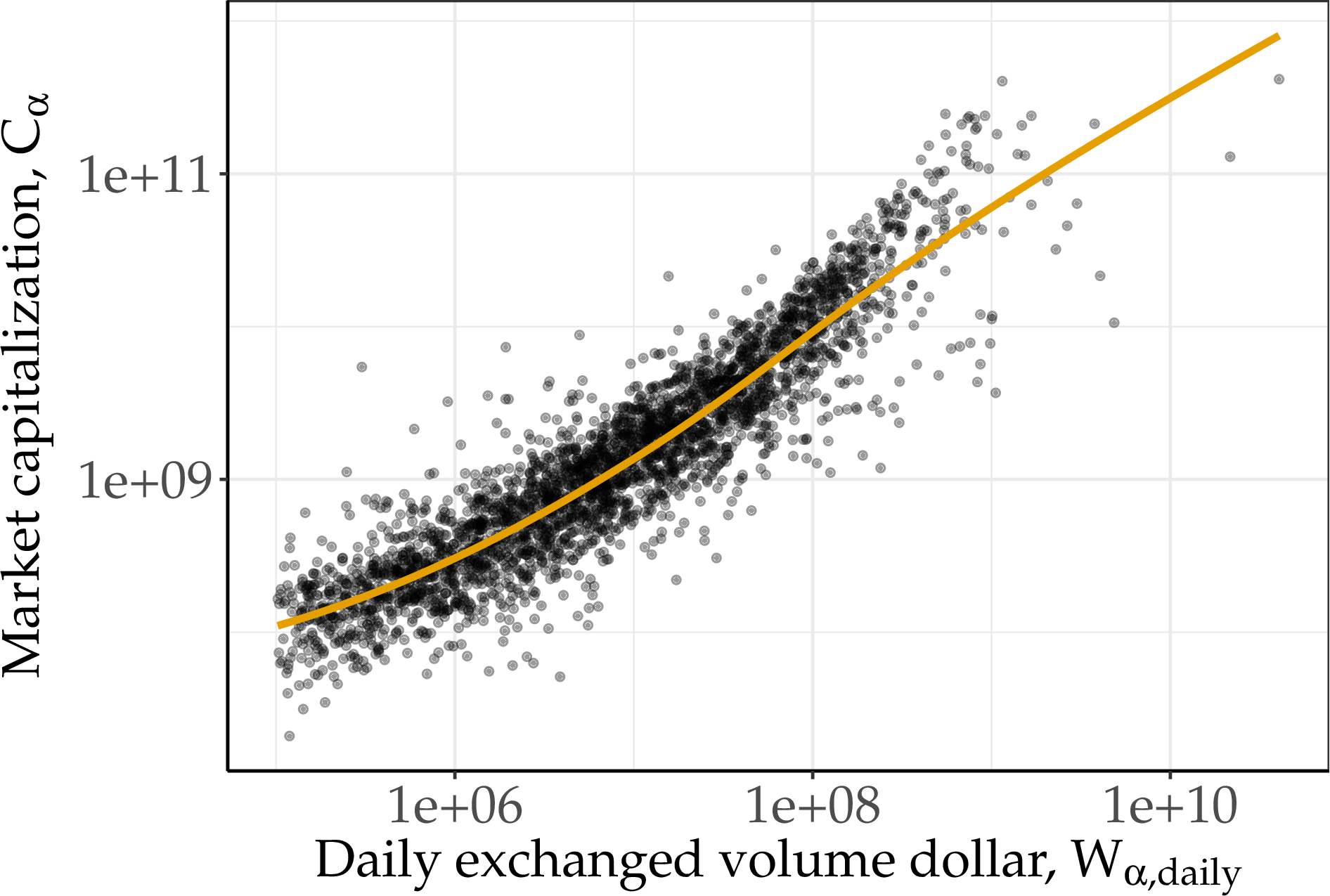}
\par\end{centering}
\caption{Market capitalization as a function of the daily exchanged volume
dollar. We find a slope close to 1 for all the dates in our database,
confirming the hypothesis that the daily exchanged volume dollar of
an asset is proportional to its market capitalization. \label{fig:ca_vs_wia}}
\end{figure}

Since the average exchanged dollar-volume of an asset is proportional
to its capitalization (Fig.~\ref{fig:ca_vs_wia}), the existence
of $f_{i}^{\text{max}}$ is a way to account for the available liquidity.

\begin{figure}
\begin{centering}
\includegraphics[scale=0.12]{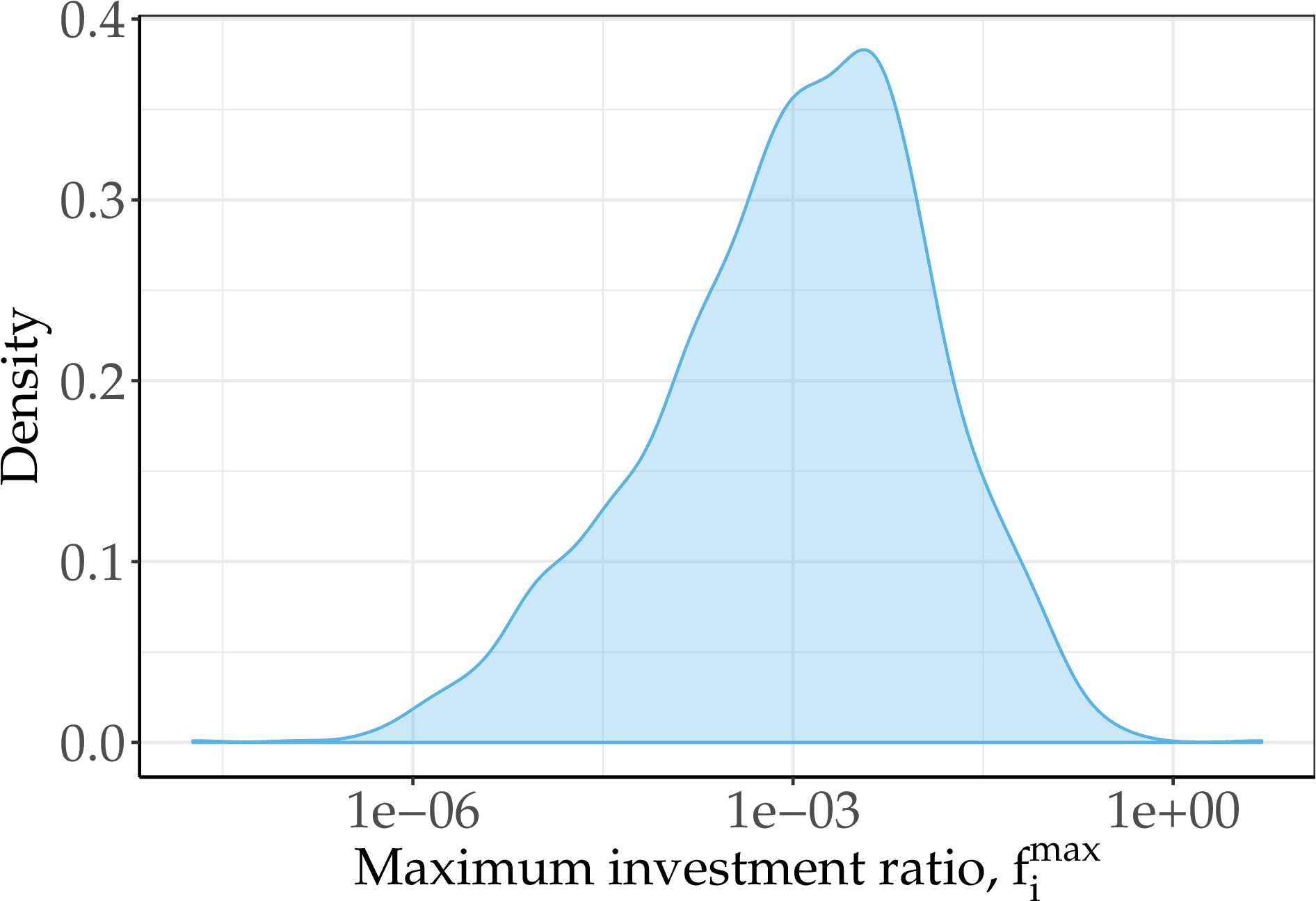}
\par\end{centering}
\caption{Empirical probability density function of $f_{i}^{\text{max}}$ for
all the funds.\label{fig:fimax_density}}
\end{figure}

Although that limit is clear for an individual fund, there is a large
range of empirical values $f_{i}^{\textrm{max}}$ Fig.~\ref{fig:fimax_density}.

\section{Simulation}

The simulation is done in a few simple steps:
\begin{enumerate}
\item Compute $n^{*}$ using the segmented model Eq.~\ref{eq:break-point}. 
\item Select a fund $i$, with a number of assets $n_{i}$.
\item If $n_{i}<n^{*}$:
\begin{enumerate}
\item Compute its optimal portfolio value using Eq.~\ref{eq:Wi_vs_ni}.
The fund will invest $\frac{W_{i}^{\text{opt}}}{n_{i}}$ for every
position.
\item Select assets randomly with a probability proportional to $C_{\alpha}$.
\end{enumerate}
\item Else if $n_{i}\geq n^{*}$:
\begin{enumerate}
\item Compute its $f_{i}^{\textrm{max}}$, so that the fund $i$ will invest
$f_{i}^{\textrm{max}}$ in $n_{i}$ assets.
\item Select assets randomly following a Beta probability distribution Fig.~\ref{fig:asset_select}
with the parameters found in Fig.~\ref{fig:beta_shape}.
\end{enumerate}
\end{enumerate}
By iterating those steps we obtain Fig.~\ref{fig:wi_vs_ni_emp}

Since the simulation outputs a portfolio for every fund, we can directly
infer the number of investors $m_{\alpha}$ of every security.

\end{document}